\documentclass[referee]{aa-mpa}
\usepackage{graphicx}
\makeatletter 
\let\@internalcite\cite
\def\cite{\@ifstar{\citeyear}{\citefull}}
\def\citefull{\def\astroncite##1##2{##1 ##2}\@internalcite}
\def\citeyear{\def\astroncite##1##2{##2}\@internalcite}
\def\citeau{\def\astroncite##1##2{##1}\@internalcite}
\def\citen{\def\astroncite##1##2{##1 (##2)}\@internalcite}
\def\possesivcite{\def\astroncite##1##2{##1's (##2)}\@internalcite}
\def\@citex[#1]#2{\if@filesw\immediate\write\@auxout{\string\citation{#2}}\fi
  \def\@citea{}\@cite{\@for\@citeb:=#2\do
    {\@citea\def\@citea{; }\@ifundefined
       {b@\@citeb}{{\bf ?}\@warning
       {Citation `\@citeb' on page \thepage \space undefined}}%
{\csname b@\@citeb\endcsname}}}{#1}}
\def\@cite#1#2{#1\if@tempswa , #2\fi}
\def\@biblabel#1{}
\makeatother

\begin{document}

\thesaurus{06(08.01.1; 08.05.3; 08.09.3; 10.07.2)}

\title{Evolution and surface abundances of red giants experiencing
deep mixing\thanks{Accepted for publication in Astron.\ \& Astrophys.}}

\author{A.~Weiss\inst{1} \and P.A.~Denissenkov\inst{1,2} \and
C.~Charbonnel\inst{3}}

\institute{Max-Planck-Institut f\"ur Astrophysik,
           Karl-Schwarzschild-Str.~1, 85748 Garching,
           Federal Republic of Germany
           \and
	   Astronomical Institute of the St. Petersburg University,
	   Bibliotechnaja Pl.~2, Petrodvorets, 198904~St.\,Petersburg,
	   Russia
           \and
           Observatoire Midi-Pyr{\'e}n{\'e}es, 14 Avenue Edouard
           Belin, 31400 Toulouse, France
}

\offprints{A.~Weiss; (e-mail: weiss@mpa-garching.mpg.de)}

\authorrunning{Weiss, Denissenkov, Charbonnel}
\titlerunning{Deep mixing in red giants}

\maketitle

\begin{abstract}
We have calculated the evolution of low metallicity red giant stars
under the assumption of deep mixing between the convective envelope
and the hydrogen burning shell.  We find that the extent of the
observed abundance anomalies, and in particular the universal O-Na
anticorrelation, can be totally explained by mixing which does not
lead to significant helium enrichment of the envelope. On the other
hand, models with extremely deep mixing and strong helium enrichment
predict anomalies of sodium and oxygen, which are much larger than the
observed ones. This latter result depends solely on the
nucleosynthesis inside the hydrogen burning shell, but not on the
details of the mixing descriptions. These, however, influence the
evolution of surface abundances with brightness, which we compare with
the limited observational material available.  Our models allow,
nevertheless, to infer details on the depth and speed of the mixing
process in several clusters. Models with strong helium enrichment
evolve to high luminosities and show an increased mass loss. However,
under peculiar assumptions, red giants reach very high luminosities
even without extreme helium mixing. Due to the consequently increased
mass loss, such models could be candidates for blue horizontal branch
stars, and, at the same time, would be consistent with the
observed abundance anomalies.

\end{abstract}

\newpage

\section{Introduction}

The observed anomalies in CNO-, NeNa- and MgAl- elements in globular
cluster red giants (see \cite{kra:94} and \cite{daco:98} for reviews)
are unexplained in canonical low-mass star evolution theory and
indicate effects beyond the standard picture. At least for anomalies
in those isotopes participating in the CNO-cycle models relying on the
assumption of an additional, non-standard mixing process inside the
stars have been presented, which explain convincingly the
observations, including the evolution of the carbon abundance along
the RGB, i.e.\ with time (see \cite{cchar:95}; \cite{dw:96};
\cite{csb:98}).
This mixing is supposed to set in after the hydrogen burning shell has
reached the composition discontinuity left behind by the first
dredge-up on the red giant branch (RGB), i.e., after the so-called RGB
bump (e.g. \cite{sm:79}; \cite{cchar:95}; \cite{cbw:98}) where the
molecular weight barrier between hydrogen burning shell and envelope
is at a minimum.  It is usually described in terms of a diffusion
process of certain efficiency and penetration depth.  The CNO
anomalies and their correlations with brightness and metallicity
(which are expected qualitatively from nucleosynthesis arguments in
standard models, as discussed by \cite{csb:98}) are thus explained by
a purely evolutionary picture (\cite{st:92}; \cite{dw:96};
\cite{bs:99b}).  The physical origin of the mixing process is believed
to be found in differential rotation of the star and the parameters
used in some of the presently available calculations have been derived
from existing theories (e.g.\ \cite{zahn:92}; \cite{mz:98}), which
are, however, far from being complete.

A similar situation holds for oxygen and sodium, which are found to be
anti-correlated in globular cluster red giants (\cite{ksn:93}).
\citen{dd:90} and \citen{lhs:93} showed that this could result from
the mixing of elements participating in the ONeNa-cycle, which
operates at higher temperatures than the CNO-cycle and therefore
requires deeper mixing.  \citen{dw:96} demonstrated how all anomalies
of the mentioned elements known at that time can be explained by the
deep mixing scenario. Their calculations were done by using canonical
red giant models, which evolved along the RGB without any mixing, and
performing the mixing and nuclear reactions in a post-processing
way. For this approach to be correct it is necessary that the
evolution of the background models is not affected by the mixing
process. In fact, the mixing necessary to reproduce the observed
anomalies was always so shallow that only very small amounts of
hydrogen/helium were mixed between envelope and shell, even in the
case of the O-Na-anti-correlation. This was taken as sufficient
justification of the underlying basic assumption. Very similar
conclusions were obtained by \citen{csb:98}.

Within this approach, one cannot investigate the possibility or
necessity for even deeper mixing, which would affect the
hydrogen/helium structure of the models.  \citen{swei:97} has renewed
the interest in such deep mixing by connecting the problem of
horizontal branch morphology with that of observed anomalies, as
previously suggested by \citen{lh:95}.  If the mixing leads to severe
helium enhancement in the envelope, increased luminosities and stellar
winds result, such that the star will populate the blue horizontal
branch (HB), while it remains a red HB star without the additional
mixing.

While \citen{dw:96} did not investigate the effect of helium
transport, \citen{swei:97} did not follow the evolution of the
participating isotopes to compare with observations. In the present
paper, we therefore attempt to close this gap by computing full
evolutionary sequences which include deep diffusive mixing and by
investigating abundance anomalies using these self-consistent models
as background models. In particular, we want to answer the question
{\em how much helium enrichment of the envelope is necessary or
allowed to achieve or to keep consistency with observations.}

In Sect.~2 we will discuss the nucleosynthesis aspects of the problem
and review the observational status of the global O-Na anticorrelation, 
which is a powerful tracer of the transport processes in the red giants.
In Sect.~3 we will present and discuss the evolutionary models. 
After that, the predictions for the abundances based on our mixed models and 
post-processing nucleosynthesis will follow, before the conclusions close 
the paper.

\section{The O-Na anticorrelation: 
observational status and nucleosynthesis arguments}

\begin{figure}
\centerline{\includegraphics[scale=0.40,draft=false]{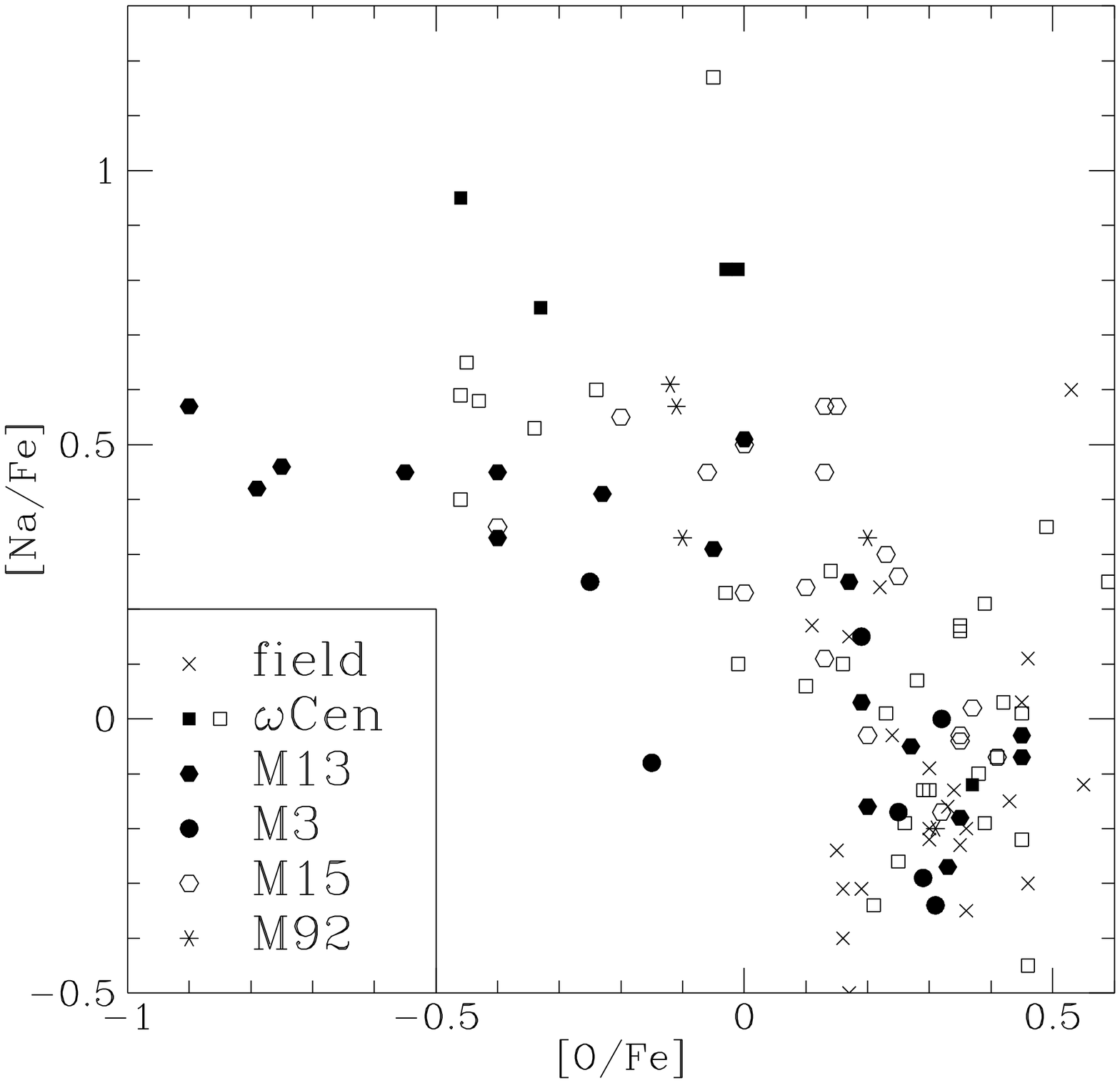}}
\caption[]{Observed [Na/Fe] vs.\ [O/Fe] for galactic globular
cluster and field giants. References for the data are:
Field: \citen{shet:96a}, \citen{shet:96b},
M3: \citen{ksl:92}, M13: \citen{kss:97}, M15: \citen{sks:97},
M92: \citen{shet:96a}, $\omega$~Cen: \citen{Norris95}. 
In $\omega$~Cen, black and white squares correspond to
stars with [Fe/H] higher and lower than -1, respectively}
\protect\label{f:obsano}
\end{figure}

Among the chemical anomalies observed in red giant atmospheres over
the past two decades, the variations in oxygen and sodium have a
special status.  Indeed, the so-called global oxygen-sodium
anticorrelation appears to be a common feature to all the globular
clusters in a wide range of metallicity 
(-2.5$\leq$[Fe/H]$\leq$-1), for which
detailed abundance analysis of the brightest giants have been carried
out (\cite{go:89}; \cite{sneden91}; \cite{drake92}; \cite{bw:92};
\cite{ksn:93}; \cite{Armosky94}; \cite{mpgc:96}; \cite{Pilachowski96};
\cite{kss:97}; \cite{ksssf:98}).  This pattern shows up in ``normal"
monometallic clusters (M3, M4, M5, M10, M13, M15, M92, NGC 7002) as
well as in the multi-metallicity cluster $\omega$ Cen
(\cite{Paltoglou89}; \cite{Norris95}). Fig.~\ref{f:obsano} summarizes
the present observational status concerning the O-Na-anticorrelation.
But more importantly, this feature, and the dependence of the Na
enhancement and O depletion on the red giant evolutionary state
(\cite{Pilachowski96}; \cite{kss:97}) can be explained
straightforwardly in the deep mixing scenario (\cite{dw:96}).

\begin{figure}[ht]
\centerline{\includegraphics[scale=0.45,draft=false,bb=60 220 580 750]{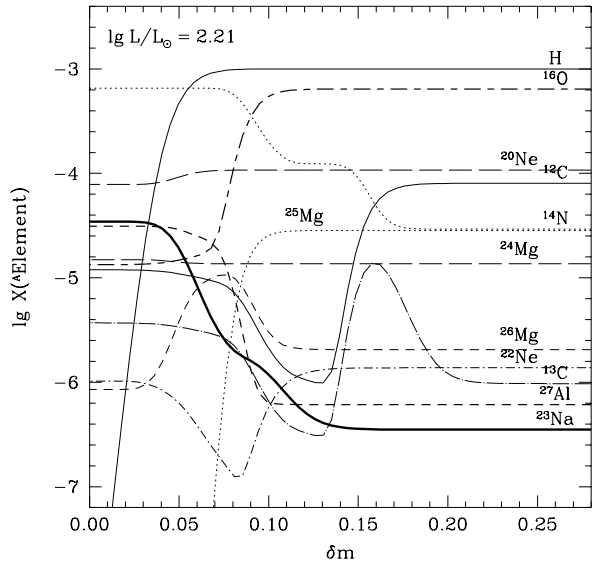}}
\caption[]{Abundances of some isotopes participating in the
         \mbox{CNO-,} NeNa- and MgAl-cycle as functions of the
         relative mass coordinate ($\delta m = 0$ at the bottom of the
         hydrogen burning shell and $\delta m = 1$ at the bottom of
         the convective envelope) in the red giant model from which detailed
        nucleosynthesis calculations with extra mixing
         start. $^{23}$Na is emphasized by a thick solid line. Also
         shown is the hydrogen abundance profile, normalized, however,
         to $10^{-3}$ to accomodate it on the same scale}
\protect\label{f:abprof}
\end{figure}

Regarding the Mg and Al anomalies, the situation is very different,
and the observed Mg-Al anticorrelation (\cite{shet:96a}) requires a
combination of the deep mixing and primordial scenarios
(\cite{dww:97}; \cite{ddn:98}).  {\em Ad hoc} assumptions would be
needed to obtain the observed $^{24}$Mg depletion within the low mass
Al-rich giants themselves (\cite{shet:96b}).  Since a low energy
resonance in the $^{24}$Mg(p,$\gamma)^{25}$Al reaction remains
undetected (\cite{Angulo99}) or even can be excluded (\cite{Pow:99}),
``exotic" models are required to episodically increase the temperature
of the hydrogen-burning shell up to values as high as $\sim$ 70-85 MK
(while canonical models reach a maximum temperature of only 55 MK) in
order to deplete Mg at the expense of $^{24}$Mg (\cite{lhz:97};
\cite{zl:97}; \cite{fmk:99}) inside the low mass giants as apparently
being the consequence of the results by \citeau{shet:96a}
(\citeyear{shet:96a} and \citeyear{shet:96b}) for M13. Lately,
however, \citen{isk:99} found that Mg and Al abundance variations in
M4 can be explained completely by the idea that the Al-enhancement is
due to a destruction of the Mg-isotopes $^{25}{\rm Mg}$ and $^{26}{\rm
Mg}$ (cf.\ Fig.~\ref{f:abprof}).  Even in this case, a significant
increase of the initial abundance of $^{25}$Mg is required
(\cite{ddn:98}). Since the Mg-Al anticorrelation cannot be explained
by the deep mixing scenario alone, we do not consider it further in
this paper. In fact, all explanations brought forward up to now being
rather exotic and complicated, one should also wait for additional
observational support for its existence (as well as for the isotopic
ratios) and relation to other stellar properties before advocating any
explanations.
 
Last but not least, the morphology of the global O-Na anticorrelation
(Fig.~\ref{f:obsano}) bears crucial clues on the mixing process.
First, its extension to very low oxygen abundances ([O/Fe] $\leq
-0.45$) is entirely due to the contribution of M13 red giants.  It is
this ``second parameter'' globular cluster that has the fastest
rotating blue horizontal branch (HB) stars.  \citen{prc:95} found six
stars having $v\sin i \geq 30\;{\rm km}\,{\rm s}^{-1}$. In the same
paper red HB stars in M3 (the cluster forming a classical ``second
parameter''-pair with M13) have been found to possess smaller
projected rotational velocities, from 2 to 20~${\rm km}\,{\rm
s}^{-1}$. This indicates a relation between rotation and mixing.

Concerning this extension along the horizontal axis, our experience in
modelling the global anticorrelation shows that it depends primarily
on the mixing rate $D_{\rm mix}$ or, more precisely, on the product
``mixing rate $\times$ mixing time''.  We will come back to these
arguments in Sect.~4, where our nucleosynthesis predictions in deeply
mixed stars will be compared to the observations of [O/Fe] versus
[Na/Fe].

Secondly, the extension of the global anticorrelation along the
vertical axis ($-0.4 \leq $ [Na/Fe] $ \leq 0.6$ in all clusters except
$\omega$ Cen) tells us mostly about the depth of the additional
mixing. In \citeau{dw:96} (\citeyear{dw:96}, Fig.~4) and in
\citeau{ddn:98} (\citeyear{ddn:98}, Fig.~1) it has been shown that on
approaching the hydrogen burning shell the Na abundance displays two
successive rises (see Fig.~\ref{f:abprof}). The first rise results
from the reaction $^{22}$Ne(p,$\gamma$)$^{23}$Na, whereas the deeper
one is produced in the NeNa-cycle by the partial consumption of
$^{20}$Ne, which is much more abundant than both $^{22}$Ne and
$^{23}$Na. The size of the vertical extension of the global
anticorrelation implies that additional mixing, whatever it is, does
not penetrate the second step (rise) in the Na abundance profile where
H starts to decrease.  Otherwise, observed [Na/Fe] values would be
much larger than they are\footnote{Actually some metal-rich giants in
$\omega$ Cen show extremely high [Na/Fe] (up to 1 dex), probably
indicating a penetration of the mixing to the second rise of Na. This
cluster is the only one where some giants exhibit surface enrichment
of Na produced from both $^{20}$Ne and $^{22}$Ne. It also has one of
the bluest horizontal branches (\cite{wor:94}).}.  \citen{csb:98} have
investigated the dependence of the abundance profiles on metallicity
(and mass). Their results confirm our arguments completely. Only for
stars of near-solar metallicity very deep mixing with significant
helium enrichment but without strong sodium enhancement could be
possible (see also Fig.~5 of \cite{dw:96}). However, the clusters
under discussion (e.g.\ M13) are metal-poor.  Let us note in
Fig.~\ref{f:obsano} the different behavior of field stars
(\cite{shet:96a}, \cite{shet:96b}). In this population, the
O-Na-anticorrelation is not present (\cite{gsc:00}). This indicates
that the deep mixing does not penetrate the region where ON-burning
occurs, and may reveal possible environmental effects on its
efficiency.

To prepare Fig.~\ref{f:abprof} we applied a nucleosynthesis code to a
red giant model with surface luminosity $\log L/L_{\odot} = 2.21$ from
which our nucleosynthesis with additional deep mixing calculations
started (Sect.~3).  The considerable growth of the $^{27}$Al abundance
with depth is due to our ``non-standard'' assumption of an enhanced
initial abundance of the $^{25}{\rm Mg}$ isotope ($[^{25}{\rm Mg/Fe}]
= 1.1$) and of the thousandfold enhanced rate of the reaction
$^{26}{\rm Al}^{\rm g}(p,\gamma)^{27}{\rm Si}$ (for details see
\cite{ddn:98}).  These ad hoc modifications were needed to explain the
observed Al enhancements (\cite{ddn:98}), but have no influence on the
results of the present paper.  {From} Fig.~\ref{f:abprof}, we see
immediately that if extra mixing penetrates down to layers, say, at
$\delta m \approx 0.06$, this will result in an enrichment of the red
giant's envelope in N, Na and Al and in its impoverishment in C, O,
and $^{25}{\rm Mg}$.  Again, $^{24}{\rm Mg}$ remains unchanged due to
the relatively low temperatures reached in such a star.  These results
were recently confirmed by \citen{plcf99} in models using the reaction
rates recommended by NACRE (\cite{Angulo99}).

\section{Red giant evolution with deep mixing}

To produce background models for the nucleosynthesis post-processing
we have evolved stellar models under the assumption of additional deep
mixing after the RGB bump. All sequences were started at the same
initial model, which consisted of an $0.8\,M_\odot$ star 
of initial composition $Y=0.25$ and $Z=0.0003$, and which had been evolved
(canonically) up to the luminosity of the bump, i.e., $\log L/L_\odot
= 2.21$ (at this point, its mass is $0.798\,M_\odot$). 
The envelope helium content has increased to 0.256 (in mass
fraction) due to the first dredge-up. The input physics of the
Garching stellar evolution code (used here) is up-to-date (for a
summary, see \cite{ddn:98}), but atomic diffusion has not been
included in the computations.

The additional deep mixing between the convective envelope and some
point inside the hydrogen shell has been implemented in the same general
line as in our previous papers on this subject, that
is, as a diffusive process with parameterized values for the
diffusive constant $D_{\rm mix}$, which is the same for all elements,
and for the penetration depth. 
The values used for $D_{\rm mix}$ are guided by the results of our
earlier papers, and agree with estimates based on rotationally induced
mixing theories. We refer the reader to \citen{dw:96} for details of
this approach.  We use the normalized mass coordinate $\delta m$
introduced therein, which is 0 at the bottom of the hydrogen shell
(usually, where $X=10^{-4}$) and 1 at the bottom of the convective
envelope.  The choice of this mass coordinate allows accurate
interpolation between a small number of background models in the
nucleosynthesis calculations (see \cite{csb:98} for a similar
approach).  The shell, in this coordinate, is located below $\delta m
\approx 0.10$. The depth, down to which the diffusive mixing should
occur, we denote as $\delta m_{\rm mix}$.  Obviously, due to the lack
of solid theories, one could also choose, for example, purely
geometrical scales to define the penetration depth (\cite{bs:99a}).
Contrary to our earlier papers, the criterion for penetration is not
determined by a fixed value for $\delta m_{\rm mix}$ chosen before the
calculations, but is related to the decrease in hydrogen content
within the shell (relative to the surface or convective envelope
abundance $X_{\rm env}$), expressed as a free parameter $\triangle X$.
We have investigated several different prescriptions for the
penetration criteria and found a great sensitivity of the mixing on
these prescriptions, which are
\begin{enumerate}	
\item find that $\delta m_{\rm mix}$ in the initial model, where
$X=X_{\rm env}-\triangle X$, and mix to the same $\delta m_{\rm mix}$
in all subsequent models;
\item as 1., but $\delta m=0$ is defined as the point where $X=X_{\rm
env}/2$ (instead of $X=10^{-4}$);
\item as 1., but the diffusion constant $D_{\rm mix}$ is decreasing
exponentially from the maximum value $D_0$ for $\delta m > 0.10$ to
$D_{\rm mix}\approx 5\cdot 10^{-5} D_0$ at $\delta m_{\rm mix}$
\item always mix to the point, where $X=X_{\rm env}-\triangle X$
\end{enumerate}
Except for method 3, these schemes were chosen to be as simple as
possible and to be similar to the one by \citen{swei:97}.

\begin{figure}[ht]
\centerline{\includegraphics[scale=0.65,draft=false]{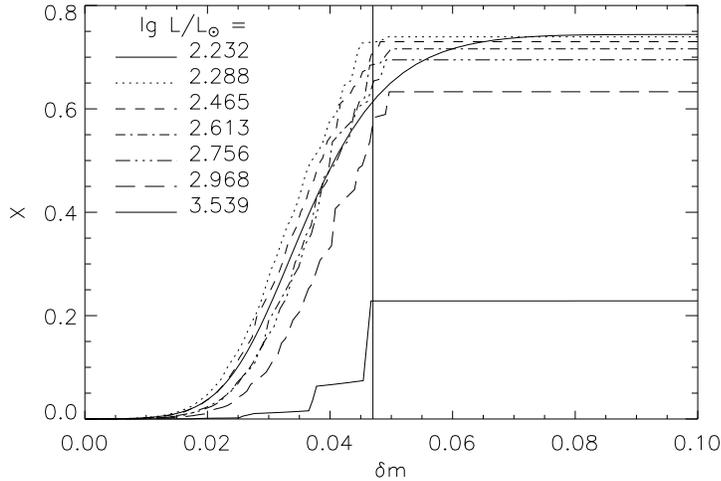}}
\caption[]{Hydrogen profiles in the normalized $\delta m$ coordinate
in models with deep mixing during the evolution along the RGB. Mixing
parameters were $D_{\rm mix}=10^9\,{\rm cm}^2\,{\rm s}^{-1}$;
$\triangle X=0.20$, method 1.  The vertical line is the mixing depth
$\delta m_{\rm mix}$ as defined in the first model. See text for more
explanations}
\protect\label{f:Xdmp24}
\end{figure}

Keeping the relative mass coordinate fixed, up to which mixing should
occur, implies that changes in the hydrogen profile in the shell
(usually steepening in the course of evolution) or in the extend of
the convective envelope influence the mixing. As an illustration of
the ``movement'' of the profile in this coordinate we display in
Fig.~\ref{f:Xdmp24} an example ($D_{\rm mix}=10^9\,{\rm cm}^2\,{\rm
s}^{-1}$; $\triangle X=0.20$; method 1). The smooth solid line is the
initial model. The next model (at $\log L/L_\odot=2.288$) is the
left-most line. From there, the model profiles shift to the right
again. The solid line with clearly reduced hydrogen abundance
throughout the envelope corresponds to a model close to the RGB-tip
($\log L/L_\odot=3.539$). In this phase the burning time at the bottom
of the mixed region becomes short enough to lead to ``bottom-burning''
of the envelope. An extended RGB-phase with luminosities drastically
increased above the canonical RGB-tip luminosity of $\log L_{\rm
tip}/L_\odot=3.33$ is the result; the final value being $\log L_{\rm
tip}/L_\odot=3.8$. Since mass loss (\cite{r:75}, with $\eta=0.3$) was
taken into account, the total mass at the tip decreases in such cases
below $0.6\, M_\odot$. $\delta m_{\rm mix}$ is indicated by the
vertical line and is $0.047$.
	
Using method 2 instead, $\delta m =0$ is defined as the point where
the abundance of hydrogen has dropped to half the surface value of the
same model ($\delta m_{\rm mix} = 0.0066$ in this case). This
prevents, obviously, any penetration to shell regions with lower
hydrogen content. As in the previous case, the bottom of the envelope
is burnt at the end of the RGB evolution. However, as soon as the
hydrogen abundance approaches $X_{\rm env}/2$ (guaranteed as long as
mixing is not quasi-instantaneous), the mixing criterion inhibits
further mixing. For this reason $X$ remains constant for $\delta m <
0$; in the final models $X_{\rm env} = 0.6402$ and the hydrogen
profile always has a finite step in the shell. In this case, the
luminosity rises to $\log L_{\rm tip}/L_\odot=3.53$.

\begin{figure}[ht]
\centerline{\includegraphics[scale=0.65,draft=false]{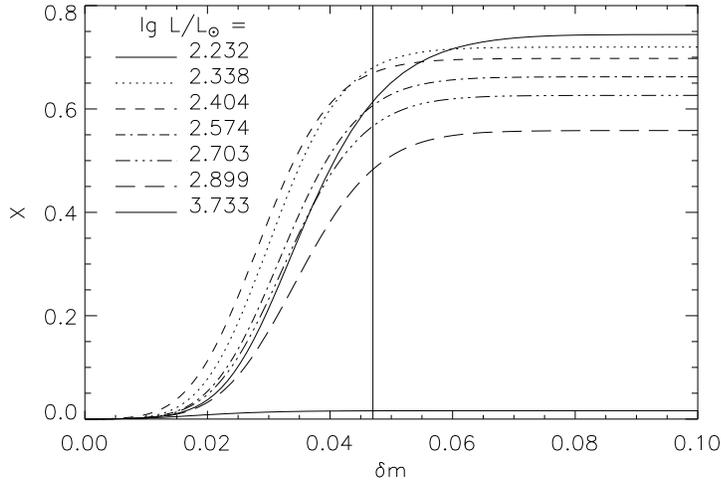}}
\caption[]{As Fig.~\ref{f:Xdmp24}, but for an exponentially declining
 diffusion coefficient $D_{\rm mix}$ (Eq.~\ref{e:Dexpo}). The results
 shown are from case D 
 in Tab.~1}
\protect\label{f:Xdme1}
\end{figure}

The steps visible in the chemical profiles of Fig.~\ref{f:Xdmp24} are
due to the diffusion criterion applied to the numerical grid, because
no interpolation to the exact value of $\delta m_{\rm mix}$ had been
done. We verified that the results do not depend on the grid
resolution, which was increased by a factor of 10 in the shell in part
of the calculations. Only the steps got smaller and more numerous. To
avoid such steps, we introduced a varying diffusion constant (method
3) motivated by recent results of \citen{dt:00}.  They have proposed a
physical mechanism for extra mixing in red giants which quantitatively
interprets all the known star-to-star abundance variations in globular
clusters.  This is Zahn's mechanism (\cite{zahn:92}; \cite{mz:98})
which considers extra mixing in a radiative zone of a rotating star as
a result of the joint operation of meridional circulation and
turbulent diffusion.  This process was already advocated by
\citen{cchar:95} to explain the low carbon isotopic ratios and lithium
abundances in field Population II giants and to lower the $^3$He
yields by low mass stars.  Denissenkov \& Tout report that the mixing
rate does not vanish abruptly at a particular depth but instead it
dies out gradually on a rather short depth range approximately between
$\delta m = 0.10$ and $\delta m = 0.06\sim 0.07$.  This explains the
following choice of an exponential decline approach for $D_{\rm mix}$:
\begin{eqnarray}
D_{\rm mix} & = & D_0; \> \delta m > \delta m_0 \nonumber \\
 & = & D_0 \exp \left[c_D \left({\delta m_0 - \delta m \over \delta m_{\rm mix} -
\delta m_0}\right)\right]; \> 0\le \delta m \le\delta m_0
\label{e:Dexpo}
\end{eqnarray}

\noindent where $\delta m_0=0.10$ was used for the beginning of the
decline and $\delta m_{\rm mix}$ is the mixing depth coordinate as
defined in method 1. Using $c_D=10$ ensures that $D_{\rm mix}(\delta
m_{\rm mix}) \approx 5\cdot10^{-5} D_0$.  The resulting evolution
(Fig.~\ref{f:Xdme1}) is similar to that of the case shown in
Fig.~\ref{f:Xdmp24}, but the profiles are smooth; mixing parameters
are identical.

We finally note that method 4, applied straightforwardly, leads to a
complete burning of the entire envelope for $D_{\rm mix} \ge 5\cdot
10^8$ and/or $\triangle X \ge 0.10$. However, this we consider to be
an artefact, which is easy to understand: Since due to the mixing the
hydrogen abundance in the envelope is reduced, the critical point down
to which mixing should occur, is moving inwards. Therefore layers of
even lower hydrogen content are mixed with the envelope, and the
critical point moves to even deeper regions. Only at the point where
the burning time-scale is shorter than the mixing time-scale and
therefore the surface hydrogen content no longer is able to adjust to
the burning, this process is stopped. A way out of this situation is
to ensure that the mixing does not lead to a sharp step in the shell's
hydrogen profile. This way, the critical point can be kept within the
mixed layers. While the mixing procedure described in \citen{swei:97}
appears to follow the straightforward approach, Sweigart (private
communication, 1999) in fact used a more complicated method to keep
the hydrogen profile even in the presence of mixing. Our method 3
qualitatively has the same effect.

To summarize, method 4 appears to be unphysical; method 3 is based on
the physical picture by \citen{dt:00}, but makes additional parameters
necessary. Methods 1 and 2 are the most straightforward choices
leading to composition profiles similar to method 3; method 2 differs
from 1 in that it avoids complete mixing and burning of the envelope
during the very last phases of RGB evolution.

\begin{table}
\caption{Parameters of calculations presented in
Figs.~\ref{f:hrdc}--\ref{f:yt} and final values of the stellar mass
($M_{\rm f}/M_{\odot}$) and luminosity ($\log L_{\rm f}/L_\odot$), and
of the helium mass fraction in the envelope ($Y_{\rm env}$).  The
final luminosity $L_{\rm f}$ is always very close or identical to the
RGB-tip luminosity $L_{\rm tip}$.  The mass loss in sequences B and C'
has been reduced by a factor 20 compared to the other ones. The
penetration depths, expressed in the normalized mass coordinate
$\delta m$ corresponding to the three $\triangle X$ values are 0.060
(0.05), 0.054 (0.10), and 0.047 (0.20). $D_{\rm mix}$ is in units of
${\rm cm}^2\,{\rm s}^{-1}$}
\protect\label{t:cases}
\begin{tabular}{l|rr|c|llc}
\hline
case & $D_{\rm mix}$& $\triangle X$ &
method & $M_{\rm f}$ & $Y_{\rm env}$ & $\log L_{\rm f}$ \\
\hline
S & 0.0 & --- & --- & 0.689 & 0.256 & 3.31\\
A & $5\cdot10^8$ & 0.05 & 3 & 0.586 & 0.270 & 3.79\\
B & $5\cdot10^8$ & 0.10 & 2 & 0.792 & 0.284 & 3.30\\
C & $10^9$ & 0.20 & 2 & 0.567 & 0.360 & 3.53\\
C'& $10^9$ & 0.20 & 2 & 0.771 & 0.360 & 3.56\\
D & $10^9$ & 0.20 & 3 & 0.562 & 0.351 & 3.77\\
\hline
\end{tabular}
\end{table}

\begin{figure}[ht]
\centerline{\includegraphics[scale=0.65,draft=false]{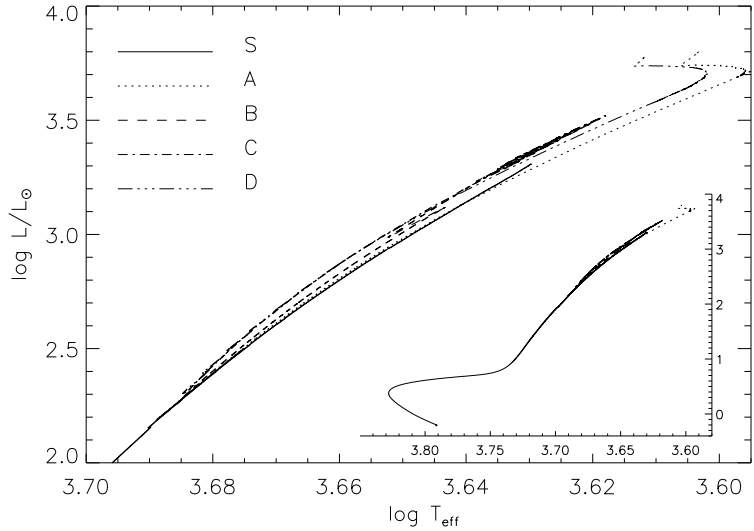}}
\caption[]{HRD of four mixed sequences of Tab.~\ref{t:cases} and
the unmixed canonical one (S) for comparison. The linetypes refer to
the cases listed in 
Tab.~\ref{t:cases} and are given in the top-left corner. The large plot
shows the upper RGB evolution only, the inset the complete evolution from
the ZAMS on.}
\protect\label{f:hrdc}
\end{figure}

\begin{figure}[ht]
\centerline{\includegraphics[scale=0.65,draft=false]{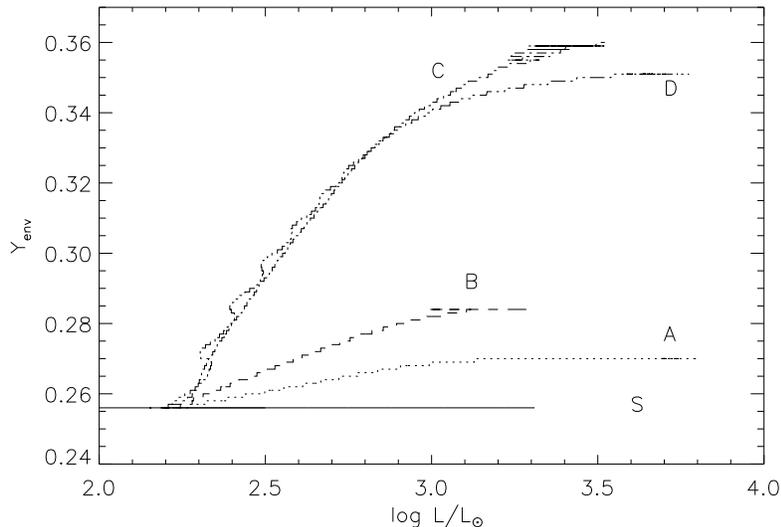}}
\caption[]{Surface helium abundance (in mass fraction)
as a function of luminosity for
the same sequences as in Fig.~\ref{f:hrdc}. A coarseness of 0.001 is
due to the limited number of digits in the output}
\protect\label{f:yl}
\end{figure}

\begin{figure}[ht]
\centerline{\includegraphics[scale=0.65,draft=false]{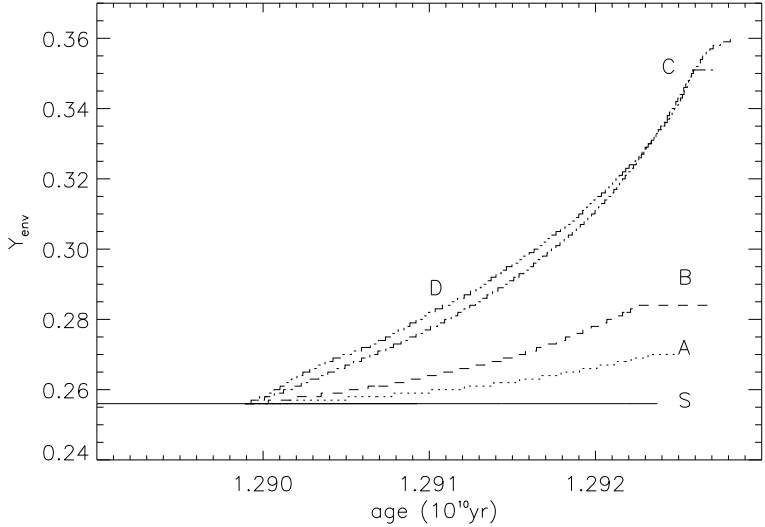}}
\caption[]{Surface helium abundance as a function of age for the same
sequences as in Fig.~\ref{f:hrdc}}
\protect\label{f:yt}
\end{figure}

Obviously the details of the mixing procedure influence the resulting
evolution to quite a significant extent. Since at present we are far
from providing a solid physical approach (which could allow, for
example, for a diffusion speed varying both in space and time), we
cannot predict the true evolution of a star experiencing deep
mixing. However, the calculations are needed only to provide
background models with varying degrees of helium mixing into the
envelope, and it is of no importance how this is achieved in detail.
We have calculated 25 different sequences, varying method and
parameters. The cases selected for Tab.~\ref{t:cases} are
representative for the range of results we obtained, which are
summarized in Figs.~\ref{f:hrdc}, \ref{f:yl}, and \ref{f:yt}.  Case D
of Tab.~\ref{t:cases} is the one also shown in Fig.~\ref{f:Xdme1}.  We
add that the amount of mass loss has no influence on the mixing
properties. Neither does a gradual switching-on of the additional
mixing during the first few models.  We have performed some comparison
calculations with a completely different code (the Toulouse-Geneva
code; \cite{cvz92}). While the results differ in details, the gross
properties are the same. The differences we ascribe to details in the
mixing procedures and the implementation of diffusion.

The effects on the evolution, displayed in Figs.~\ref{f:hrdc} --
\ref{f:yt}, are qualitatively as expected from the work by
\citen{swei:97}. The increase in the surface helium content is quite
dramatic in cases C and D (fast and very deep mixing). However, it is
not as large as in \citen{swei:97}, shown, for example, in his Fig.~2,
where for $\triangle X = 0.20$ a value of $Y\ga 0.42$ was
reached. Also, in contrast to Sweigart's result, in all our
calculations the helium enrichment of the outer envelope tends to
level off with progressing evolution. This might be ascribed again to
differences in the mixing scheme details, as we find these differences
also in the post-processing models presented in the next section
(cf. Figs.~\ref{f:naoyl2} and \ref{f:naoyl1}). We also find that the
luminosities can get extremely high (cases A and D) with high mass
loss as the consequence and a beginning turn-away from the RGB before
the He-flash sets in. The beginning of such an evolution might be
recognised, too, in Fig.~3 of \citen{sweip:97} in the case of deepest
mixing. We also note that mixing method 2 (cases B and C) results in
loops in the HRD, which depend on the occurrence of mixing
episodes. This is, for example, visible in the non-monotonic
luminosity evolution in Fig.~\ref{f:yl}.

The lifetime on the RGB is in all cases prolonged (Fig.~\ref{f:yt}) by
1 or 2~Myr, in some cases up to 4 Myr (5--10\% of the lifetime after
the bump). The evolutionary speed is influenced mainly immediately
after the onset of the additional mixing (after the bump). From the
initial model to one at $\log L/L_\odot \approx 2.5$, the time
increases from 7~Myr (case S) to about 10~Myr (A \& B) and 13~Myr (C
\& D). Thereafter, it is becoming comparable again in all cases. Such
an effect could possibly be seen in luminosity functions, but it
should be most prominent only in a limited luminosity
range. \citen{vlp:98} have argued that the luminosity function of M30
could be evidence for rapidly rotating cores of giants.

\begin{figure}[ht]
\centerline{\includegraphics[scale=0.65,draft=false]{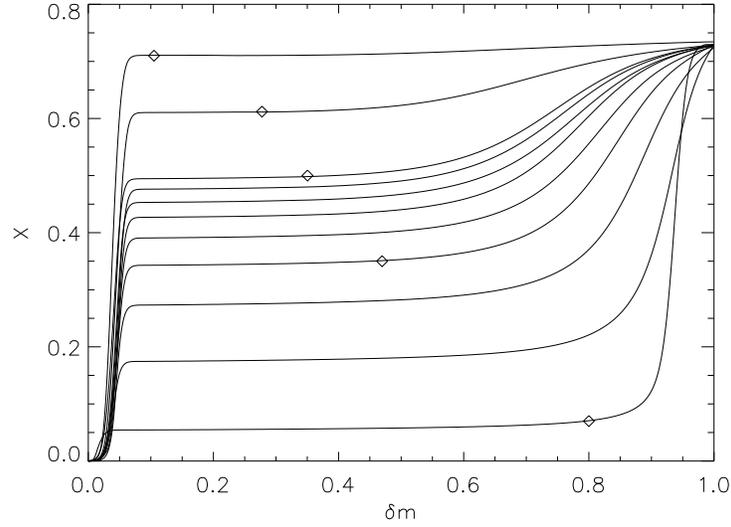}}
\caption[]{Hydrogen profiles of selected models at the tip of 
sequence A ($\triangle X=0.05$, $D_{\rm mix}=5\cdot10^8\,{\rm
cm}^2\,{\rm s}^{-1}$, mixing method 3). 
The point where energy generation by hydrogen burning exceeds
$10^3\,{\rm erg/(gs)}$ first, is indicated in several models by symbols.}
\protect\label{f:mhil}
\end{figure}

In some calculations, in particular those employing the exponentially
declining diffusive speed (method 3), an interesting effect
appeared. Although the mixing for moderate penetration depth and
mixing speed remains small, the luminosity of the models rises to
extreme values. As an example, in case A ($\triangle X=0.05$, $D_{\rm
mix}=5\cdot10^8\,{\rm cm}^2\,{\rm s}^{-1}$), $Y_{\rm env} = 0.270$ (an
enrichment of only $+0.014$) and $\log L/L_\odot = 3.79$ were reached
at the tip of the RGB (see also Fig.~\ref{f:hrdc}).  An inspection of
the models shows that part of the extended region of almost
homogeneous composition, which is achieved due to the effect of the
additional diffusion, becomes hot enough for significant hydrogen
burning. In Fig.~\ref{f:mhil} we display the H-profiles of selected
models in this phase. The first one (top line) is at $\log L/L_\odot =
2.73$ ($Y_{\rm env} = 0.265$). In this model, the point were the
energy production due to hydrogen burning exceeds $10^3\,{\rm
erg}\,{\rm g}^{-1}\,{\rm sec}^{-1}$ for the first time, is very close
to the composition step. In the next model ($\log L/L_\odot = 3.12$;
$Y_{\rm env} = 0.269$) this point has shifted to $\delta m = 0.274$
and is constantly progressing outward until it reaches $\delta m =
0.80$ in the most advanced models at the bottom of the figure ($\log
L/L_\odot = 3.68$; $Y_{\rm env} = 0.270$). This burning of the plateau
constitutes a broadening of the hydrogen shell and delivers extra
luminosities. In fact, 50\% of the total luminosity of the models with
the plateau value around $X=0.40$ are generated at $\delta m > 0.06$,
that is outside the inner composition step. Therefore, the luminosity
in excess of that of an ordinary star at the tip of the RGB (around
$\log L/L_\odot = 3.30$) can completely be ascribed to the plateau
burning (note that diffusion cannot keep the burning plateau
homogeneous with the outer regions). The effect we observe here is
probably due to our mixing description, which sets $D_{\rm mix}$ to
the maximum value outside $\delta m = 0.10$.  If our mixing
description is realistic, this would imply that one could get very
high luminosities at the RGB-tip {\em without} extreme helium mixing.
The time spent at luminosities above the standard TRGB brightness is
only $10^6$~yrs and therefore observation of such a superluminous star
is rather unlikely. Due to the extreme overluminosities, the Reimers
mass loss formula leads to stellar winds of order
$10^{-7}\,M_\odot$/yr and a final mass of $0.58\,M_\odot$
($M_c=0.517\,M_\odot$).  After the helium flash, this star will
populate the blue part of the horizontal branch.
	
\newpage

\section{Nucleosynthesis in deeply mixed stars}

The detailed nucleosynthesis calculations we present now have been
performed in a post-processing way as in our previous works
(\cite{dw:96}; \cite{ddn:98}). From an evolutionary sequence three red
giant models were selected.  The starting one was the same one as for
the full evolutionary calculations discussed in the previous section,
i.e.\ a model at the bump ($\log L/L_\odot \approx 2.2$) in which the
hydrogen burning shell had recently crossed the H-He discontinuity
left by the base of the convective envelope on the first dredge-up
phase. The final one was a model near the RGB tip ($\log L/L_\odot
\approx 3.3$), the second one having a luminosity intermediate to
those of the starting and finishing models. Distributions of $T$,
$\rho$ and $r$ with $\delta m$ in these three ``background'' models
were used for interpolations in $\log\,L$ during the nucleosynthesis
calculations. Further details about our post-processing procedure can
be found in \citen{dw:96}. The network of nuclear kinetics equations
was the smallest one of those considered in \citen{ddn:98}. It takes
into account 26 particles coupled by 30 nuclear reactions from the
pp-chains, CNO-, NeNa- and MgAl-cycle. The additional mixing is
modelled by diffusion with a constant coefficient $D_{\rm mix}$.  We
recall that we allow for mixing prescriptions and parameters in these
calculations different from those for which the background models have
been obtained.

\begin{figure}
\centerline{\includegraphics[scale=0.46,bb= 0 220 540 700,draft=false]{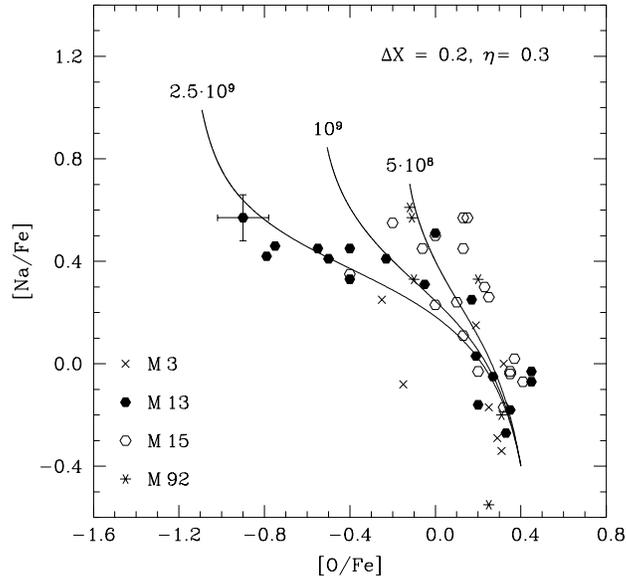}}
\caption[]{The global anticorrelation and three theoretical curves
            (labelled by the value of $D_{\rm mix}$ used in the
            corresponding calculation);
            all three curves were calculated with the penetration
            criterion 1 and $\triangle X = 0.2$; {\em unmixed} red giant
            models were used as background. Observational data
            (omitting $\omega~Cen$) are as in Fig.~\ref{f:obsano}}
\protect\label{f:naoanti2}
\end{figure}

For the comparison with observations we have preferred the ``global
anticorrelation'' of [O/Fe] versus [Na/Fe] for the reasons detailed in
Sect.~2. In Fig.~\ref{f:naoanti2} and Fig.~\ref{f:naoanti1} it is
plotted for globular clusters according to the latest observational
data.

\begin{figure}
\centerline{\includegraphics[scale=0.46,draft=false,bb= 0 220 540 700]{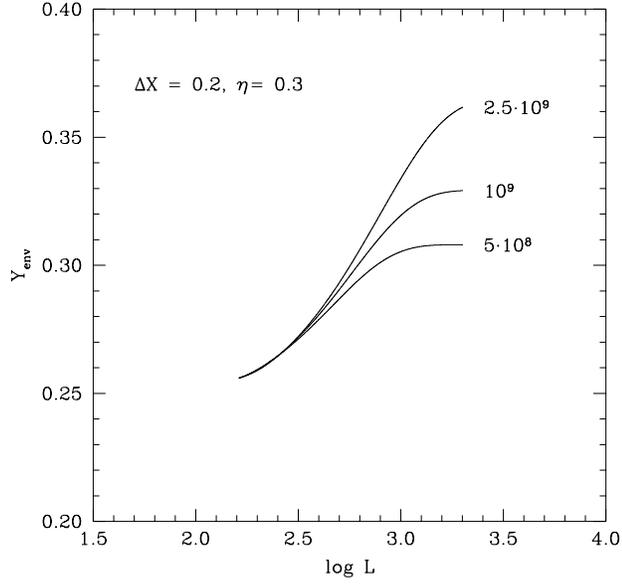}}
\caption[]{The same three curves as in Fig.~\ref{f:naoanti2} in the
$\log L$ - $Y_{\rm env}$ plane}
\protect\label{f:naoyl2}
\end{figure}

In Fig.~\ref{f:naoanti2} theoretical dependences of [Na/Fe] on [O/Fe]
obtained in the post-processing way for three values of the diffusion
coefficient are shown. This first set of calculations was performed
with {\em unmixed} background red giant models like in our previous
papers but with mass loss taken into account.  The depth of additional
mixing was determined according to the penetration criterion 1
(Sect.~2) with $\triangle X = 0.2$. The mass loss rate $\dot{M}$ was
estimated with \citen{r:75} formula in which the parameter value $\eta
= 0.3$ was adopted.  In Fig.~\ref{f:naoyl2} the resulting envelope He
abundances are shown as functions of $\log\,L/L_\odot$. The mixing
depth in the starting model which was kept constant during the
nucleosynthesis calculations was $\delta m_{\rm mix} = 0.047$. Such a
value of $\delta m_{\rm mix}$ allows some (modest) penetration of the
second Na step (see Sect.~2) by the mixing which results in an upward
steepening of the theoretical dependences of [Na/Fe] on [O/Fe] by the
end of the RGB evolution (Fig.~\ref{f:naoanti2}).  The maximum He
enrichment achieved in the envelope in this set of calculations is
$\triangle Y_{\rm env} \approx 0.10$ (Fig.~\ref{f:naoyl2}).
	
\begin{figure}
\centerline{\includegraphics[scale=0.46,bb= 0 220 540 700,draft=false]{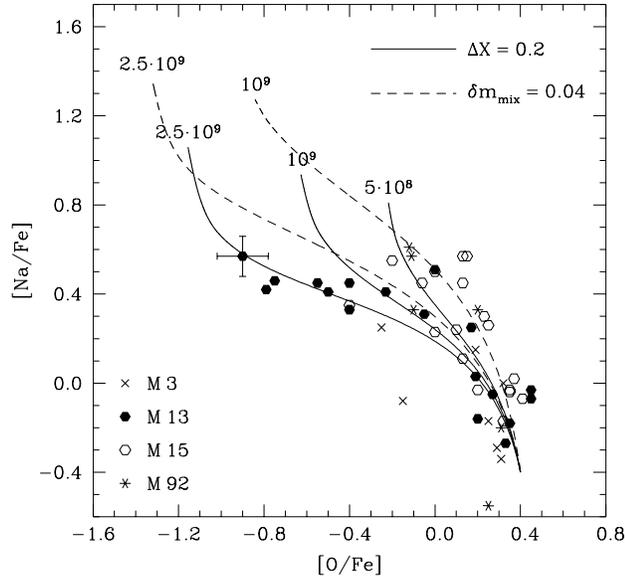}}
\caption[]{As Fig.~\ref{f:naoanti2}, but for calculations using {\em
            mixed} red giant models as background.
            The solid curves were calculated with the penetration
            criterion 1 and $\triangle X = 0.2$ and the dashed ones
            with constant mixing depth $\delta m_{\rm mix} = 0.04$}
\protect\label{f:naoanti1}
\end{figure}
	
\begin{figure}
\centerline{\includegraphics[scale=0.46,draft=false,bb= 0 220 540 700]{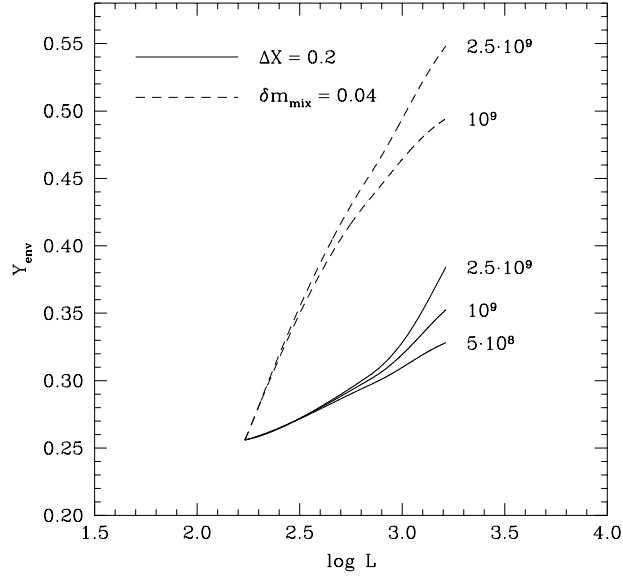}}
\caption[]{The same five curves as in Fig.~\ref{f:naoanti1} in the
$\log L$ - $Y_{\rm env}$ plane}
\protect\label{f:naoyl1}
\end{figure}

In the next nucleosynthesis calculations, the results of which are
presented in Figs.~\ref{f:naoanti1} and \ref{f:naoyl1}, we have used
{\em mixed} background models from sequence C' of Tab.~\ref{t:cases};
for this sequence the same mixing parameters as in case C
(Tab.~\ref{t:cases}) but a reduced mass loss rate has been used: the
parameter $\eta$ was divided by 20 in order to take into account a
reduction of $\dot{M}$ for low Z (\cite{maed:92}). The reduced mass
loss rate affects only the final mass, but not the helium enrichment.
These models were evolved along the RGB with effects of the He mixing
on stellar structure parameter distributions fully taken into account
in the stellar evolution code (Sect.~3).  Sequence C' was chosen
because out of the sample cases listed in Tab.~\ref{t:cases} it has
the highest degree of helium enrichment, in contrast to the previous,
unmixed, background models.  In this second set of nucleosynthesis
calculations we repeat the case of the first set
(Fig.~\ref{f:naoanti2}), i.e.\ mixing down to $\triangle X = 0.2$
(solid lines), but also add two computer runs (dashed lines) in which
the mixing was chosen to be so deep that a high enrichment in He of
the envelope was guaranteed.  We label these calculations by $\delta
m_{\rm mix} = 0.04$, which is the penetration depth needed to mix down
to $\triangle X = 0.37$.

Comparison of the results obtained in the two sets of calculations
allows to draw the following conclusions:
\begin{itemize}
\item mass loss is practically unimportant for this study (at least
within the prescriptions and variations used in the various
calculations);
\item making use of mixed background models instead of unmixed ones
does not seriously affect the theoretical dependences of [Na/Fe] on
[O/Fe] (compare the solid curves in Figs.~\ref{f:naoanti2} and
\ref{f:naoanti1}); therefore the details of the mixing prescription
used for the background models are not significant for the
nucleosynthesis results.
\item the total He enrichment of the convective envelope calculated
in the post-processing way agrees very well with the final envelope He
abundance obtained in the full evolutionary
calculations with additional deep mixing;
\item values $\triangle Y_{\rm env} > 0.15$ were obtained only in the two
computer runs with the depth $\delta m_{\rm mix} = 0.04$, but in these
cases additional mixing penetrated so deeply that it resulted in the
[Na/Fe] on [O/Fe] dependences evidently inconsistent with the observations
(dashed curves in Figs.~\ref{f:naoanti1} and \ref{f:naoyl1}).
\end{itemize}

A simple inspection of Figs.~\ref{f:naoanti2}-\ref{f:naoyl1} allows
the conclusion that {\em the global anticorrelation of [O/Fe]
vs. [Na/Fe] as a whole and especially the [Na/Fe] values in its low
oxygen abundance tail certainly rule out any hypothesis about an
increase of more than $\triangle Y_{\rm env} \approx 0.10$ in the
envelope He abundance of globular-cluster red giants.}  A physical
reason for this constraint is the above-mentioned inability of
additional mixing to penetrate the second Na abundance rise lying at
$\delta m \leq 0.06 \div 0.07$ as hinted by the observed global
anticorrelation. In the starting models $\delta m \geq 0.07$
corresponds to $\triangle X \leq 0.05$ and, therefore, the envelope He
enrichment is not expected to be much larger than $\triangle Y_{\rm
env} \approx 0.05$.

\section{Discussion}

If the O-Na-anticorrelation observed in many globular cluster red
giants is indeed due to a deep mixing process beyond the standard
effects taken into account in canonical stellar evolution theory, the
question is justified whether this deep mixing might affect
the H-He-profile as well. In this case, consequences for the red
giant evolution including phases of enhanced luminosities and mass
loss could result. We have, therefore, discussed both the
evolutionary and nucleosynthetic effects quantitatively by performing
stellar evolution calculations including deep mixing and
post-processing nucleosynthesis models (the latter
as we did in our earlier papers \cite{dw:96}; \cite{ddn:98}). 

{From} arguments depending only on nucleosynthesis we could already
infer that for temperature profiles typical of hydrogen-burning shells
mixing of appreciable amounts of helium can only be achieved if the
second Na rise is penetrated. This, however, leads to oxygen and
sodium anomalies exceeding those observed (with the exception of a few
stars in $\omega$~Cen, a multi-metallicity, untypical cluster).  In
terms of our normalized mass coordinate (defined such that $\delta m =
0$ at $X = 10^{-4}$ at the bottom of the shell) this puts an
observationally constrained limit for the maximum mixing depth of
$\delta m_{\rm mix} > 0.06 \div 0.07$.  Our complete models confirm
this argument: helium enrichment in excess of $\triangle Y_{\rm env}
\approx 0.05$ due to deep mixing can be ruled out for those stars with
Na-O-anomalies as observed in clusters such as M15, M92, M3, and even
M13 which presents one of the most extended blue horizontal branch.

\begin{figure}
\centerline{\includegraphics[scale=0.45,draft=false]{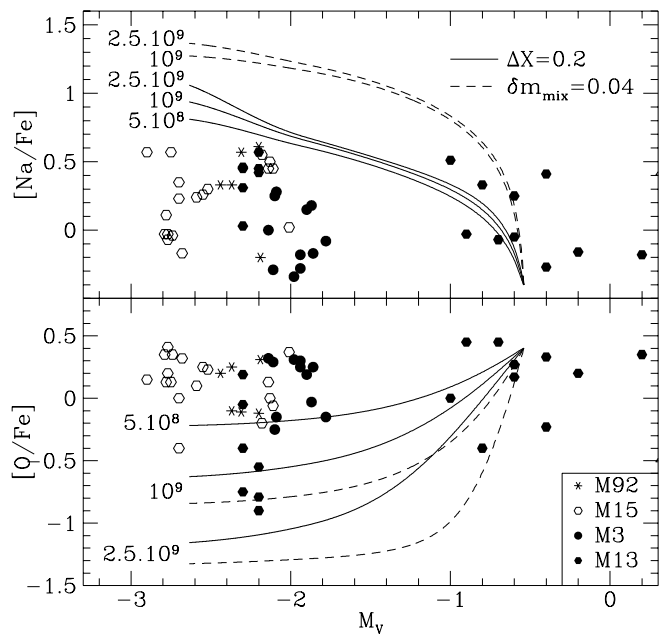}}
\caption[]{[Na/Fe] and [O/Fe] for giants in globular clusters. The
symbols identify individual clusters and have the same meaning as in
Fig.~\ref{f:obsano}. Overlaid are the five theoretical predictions
shown in Fig.~\ref{f:naoanti1}, labeled by the values of $D_{\rm
mix}$. In the lower panel the two higher values relate to two lines
each.}
\protect\label{f:naomv}
\end{figure}

The details of the surface abundance history along the RGB depend on
the details of the deep mixing process and thus on its nature which we
did not attempt to specify here.  In particular, the field-to-cluster
differences must be understood (especially the fact that field giants
do not present the O-Na anticorrelation (\cite{gsc:00}),
indicating a deeper and more efficient mixing in their globular
cluster counterparts) and seem to point out a non-negligible impact of
environmental effects on the extra mixing efficiency.

However, we can compare the observations with the histories of Na and
O abundance anomalies predicted by our simple mixing prescriptions in
order to get constraints for a solid physical model.  This is done in
Fig.~\ref{f:naomv}, which shows surface abundances for several
clusters as a function of brightness, i.e.\ progressing
evolution. Also shown is the theoretical prediction of the five
calculations displayed in Fig.~\ref{f:naoanti1}. Although the
observational data are very few (the uncertainty in the abundances is
of order 0.2 dex), and information for stars before or at the bump is
available only for M13, some effects can be recognized, nevertheless.

All clusters show the whole spread of Na abundances between the
canonical case without extra mixing (they would lie on a horizontal
line) and that as obtained from the calculations with less deep mixing
(solid lines).  As already mentioned, very low O abundances are only
found in M13 giants, and the observed spread for this element is a
signature of the mixing rate.  For both elements the abundance
anomalies are limited to values predicted by models with extra mixing
{\em not} penetrating the second $^{23}$Na rise in the hydrogen
shell. Stars with intermediate anomalies we interpret as being due to
mixing penetrating less deeply into the hydrogen shell. They could be
reproduced with properly adjusted mixing parameters.

Over the small brightness interval for which we have data (for all the
clusters except M13 only the brightest giants are accessible), no
significant abundance evolution is recognizable.  The increase in $\rm
[Na/Fe]$ at the RGB tip obtained in all calculations labeled
$\triangle X=0.20$ is not visible in the observations, which may be
taken as indication that the physical reasons (e.g.\ rotation) for the
additional mixing have lost their importance or even vanished.  Such a
possible time dependence of the extra mixing has not been taken into
account in our simple mixing prescriptions, but should result from
more physically motivated models. (We recall that in \citen{dw:96} the
carbon evolution could be reproduced with constant mixing parameters,
however.) We will therefore, in a forthcoming paper, use the model by
\citen{dt:00}, which includes temporal changes in the diffusion
constant due to angular momentum transport.

In our models, the largest abundance changes take place at the onset
of the additional mixing, i.e., early on the RGB, after the bump
(around $M_V\approx -0.5$). This might be visible in the M13 data.
The low brightness group has normal abundances, but above $M_V\approx
-0.4$ strong anomalies already appear, reminiscent of the steep
increase in the calculations mixing deeper (dashed lines). In the case
of Na no further enhancement is visible (the range of values does not
varies along the RGB), while O seems to get depleted further. In spite
of the incomplete data at hand, the abundance anomalies seem to
develop rather early and within a narrow brightness range, but then do
not increase any more. This general behaviour is consistent with our
theoretical predictions -- though not exactly reproduced -- and again
points to not too deep mixing at moderate speed. The calculations with
extreme helium mixing would predict the largest O depletion and Na
enhancement all along the RGB; the absence of such stars can therefore
not be explained with a selection effect working against the shortest
lived stars at the tip of the RGB. The fact that M13 seems to show
anomalies already before the bump has to be taken with care, because
we compare here only with one stellar model which has a metallicity
almost a factor of 10 smaller than that appropriate for M13. At that
metallicity, the bump would occur about 1~mag earlier (and our initial
model is about 0.2~mag brighter than the end of the bump).  We stress
the fact that the present observational data do not allow a detailed
comparison with the abundance evolution, but only rough qualitative
statements.

Although Fig.~\ref{f:naomv} again demonstrates how the abundance
evolution depends on the assumption about the additional mixing
process, our calculations also show that the nucleosynthesis argument
given above is valid for {\em all} mixing descriptions tested and
therefore model-independent. From this we predict that {\sl red giants
exhibiting the observed Na-O-anomalies do not have envelopes enriched
in helium by much more than $\triangle Y_{\rm env} \approx 0.05$,
which is comparable to the general uncertainty in our knowledge about
the helium abundances in such stars.}

Concerning the brightness reached during the RGB evolution, we showed
that for specific mixing prescriptions large excess luminosities can
be achieved at the end of the evolution without simultaneous mixing of
large amounts of helium. As the reason we could identify the burning
of the outer parts of the hydrogen shell where some H-He-mixing had
happened in earlier phases. These regions become hot enough for
significant hydrogen burning on such short time-scales that the deep
mixing process is not able to mix the products any longer into the
convective envelope. Such models experience very strong mass loss
under the assumption of the continuing validity of a Reimers-type
stellar wind and finish the RGB phase with total masses below
$0.6\,M_\odot$ and envelope masses of only 10\% of this. They could be
candidates for blue HB stars and would link observed abundance
anomalies, deep mixing and the second parameter problem, as suggested
by \citen{lh:95} and \citen{swei:97}. They would avoid the problem of
overproducing O-Na-anomalies as would result from the helium mixing
investigated by \citen{swei:97}.

However, we repeat our warning that the details of the evolutionary
consequences of deep mixing depend crucially on the mixing process and
history. Mixing speed and depth are both important for the amount of
helium mixed and for the result of the competing processes ``mixing''
and ``burning'' and their time-scales. The mixing depth is also linked
to the criterion for deep mixing, for which we have investigated
several simple recipes.

To conclude, we need a solid physical picture for the deep mixing
process in order to be able to investigate its effect on red giant
evolution further. Presently, we can only point out some interesting
possibilities -- such as the overluminosities -- and derive
model-independent features, such as our main conclusion that the
observed anomalies of oxygen and sodium rule out strong helium
enhancement and therefore very deep mixing.

\begin{acknowledgements}
We are grateful to A.~Sweigart for helpful discussions.  This study
was partly done while CC and PAD visited the Max-Planck-Institut f\"ur
Astrophysik in Garching. They express their gratitude to the staff for
hospitality and support. We appreciate the very careful work of
an anonymous referee, whose detailed and constructive comments helped
to improve this paper.
\end{acknowledgements}

\newpage
%

\begin{thebibliography}{50}

\bibitem[\protect\astroncite{Angulo et~al.}{1999}]{Angulo99}
Angulo C., Arnould M., Rayet M., et~al., 1999, Nucl.Phys. A 656, 3

\bibitem[\protect\astroncite{Armosky et~al.}{1994}]{Armosky94}
Armosky B.J., Sneden C., Langer G.E., Kraft R.P., 1994, AJ 108, 1364

\bibitem[\protect\astroncite{Boothroyd \& Sackmann}{1999{a}}]{bs:99b}
Boothroyd A.I., Sackmann I.J., 1999{a}, ApJ 510, 232

\bibitem[\protect\astroncite{Boothroyd \& Sackmann}{1999{b}}]{bs:99a}
Boothroyd A.I., Sackmann I.J., 1999{b}, ApJ 510, 217

\bibitem[\protect\astroncite{Brown \& Wallerstein}{1992}]{bw:92}
Brown J.A., Wallerstein G., 1992, AJ 104, 1818

\bibitem[\protect\astroncite{Cavallo et~al.}{1998}]{csb:98}
Cavallo R., Sweigart A., Bell R., 1998, ApJ 492, 575

\bibitem[\protect\astroncite{Charbonnel}{1995}]{cchar:95}
Charbonnel C., 1995, ApJL 453, L41

\bibitem[\protect\astroncite{Charbonnel et~al.}{1998}]{cbw:98}
Charbonnel C., Brown J.A., Wallerstein G., 1998, A\&A 332, 204

\bibitem[\protect\astroncite{Charbonnel et~al.}{1992}]{cvz92}
Charbonnel C., Vauclair S., Zahn J.P., 1992, A\&A 255, 191

\bibitem[\protect\astroncite{{Da~Costa}}{1998}]{daco:98}
{Da~Costa} G.S., 1998, in T.~R. Bedding, A.~J. Booth, and J. Bavis (eds.),
  Fundamental Stellar Properties: the interaction between observation and
  theory, no. 189 in IAU Symp. Kluwer, Dordrecht, p.~193

\bibitem[\protect\astroncite{Denissenkov et~al.}{1998}]{ddn:98}
Denissenkov P.A., {Da~Costa} G.S., Norris J.E., Weiss A., 1998, A\&A 333, 926

\bibitem[\protect\astroncite{Denissenkov \& Denissenkova}{1990}]{dd:90}
Denissenkov P.A., Denissenkova S.N., 1990, SvA Lett. 16, 275

\bibitem[\protect\astroncite{Denissenkov \& Tout}{2000}]{dt:00}
Denissenkov P.A., Tout C.A., 2000, MNRAS, accepted

\bibitem[\protect\astroncite{Denissenkov \& Weiss}{1996}]{dw:96}
Denissenkov P.A., Weiss A., 1996, A\&A 308, 773

\bibitem[\protect\astroncite{Denissenkov et~al.}{1997}]{dww:97}
Denissenkov P.A., Weiss A., Wagenhuber J., 1997, A\&A 320, 115

\bibitem[\protect\astroncite{Drake et~al.}{1992}]{drake92}
Drake J.J., Smith V.V., Suntzeff N.B., 1992, ApJ Letters 395, 95

\bibitem[\protect\astroncite{Fujimoto et~al.}{1999}]{fmk:99}
Fujimoto M.Y., Aikawa M., Kato K., 1999, ApJ 519, 733

\bibitem[\protect\astroncite{Gratton \& Ortolani}{1989}]{go:89}
Gratton R.G., Ortolani S., 1989, A\&A 211, 41

\bibitem[\protect\astroncite{Gratton et~al.}{2000}]{gsc:00}
Gratton R.G., Sneden C., Carretta E., Bragaglia A., 2000, A\&A 354, 169

\bibitem[\protect\astroncite{Ivans et~al.}{1999}]{isk:99}
Ivans I.I., Sneden C., Kraft R.P. et~al., 1999, AJ 118, 1273

\bibitem[\protect\astroncite{Kraft}{1994}]{kra:94}
Kraft R.P., 1994, PASP 106, 553

\bibitem[\protect\astroncite{Kraft et~al.}{1997}]{kss:97}
Kraft R.P., Sneden C., Smith G.H. et~al., 1997, AJ 113, 279

\bibitem[\protect\astroncite{Kraft et~al.}{1998}]{ksssf:98}
Kraft R.P., Sneden C., Smith G.H., Shetrone M.D., Fulbright J., 1998, AJ 115,
  1500

\bibitem[\protect\astroncite{Kraft et~al.}{1992}]{ksl:92}
Kraft R.P., Sneden C., Langer G.E., Prosser C.F., 1992, AJ 104, 645

\bibitem[\protect\astroncite{Kraft et~al.}{1993}]{ksn:93}
Kraft R.P., Sneden C., Langer G.E., Shetrone M.D., 1993, AJ 106, 1490

\bibitem[\protect\astroncite{Langer \& Hoffman}{1995}]{lh:95}
Langer G.E., Hoffman R.D., 1995, PASP 107, 1177

\bibitem[\protect\astroncite{Langer et~al.}{1993}]{lhs:93}
Langer G.E., Hoffman R.D., Sneden C., 1993, PASP 105, 301

\bibitem[\protect\astroncite{Langer et~al.}{1997}]{lhz:97}
Langer G.E., Hoffman R.D., Zaidins C.S., 1997, PASP 109, 244

\bibitem[\protect\astroncite{Maeder}{1992}]{maed:92}
Maeder A., 1992, A\&A 264, 105

\bibitem[\protect\astroncite{Maeder \& Zahn}{1998}]{mz:98}
Maeder A., Zahn J.P., 1998, A\&A 334, 1000

\bibitem[\protect\astroncite{Minniti et~al.}{1996}]{mpgc:96}
Minniti D., Peterson R.C., Geisler D., Clarin J.J., 1996, ApJ 470, 953

\bibitem[\protect\astroncite{Norris \& Costa}{1995}]{Norris95}
Norris J.E., Costa G.S.D., 1995, ApJL 441, 81

\bibitem[\protect\astroncite{Palacios et~al.}{1999}]{plcf99}
Palacios A., Leroy F., Charbonnel C., Forestini M., 1999, preprint
  astro-ph/9910289

\bibitem[\protect\astroncite{Paltoglou \& Norris}{1989}]{Paltoglou89}
Paltoglou G., Norris J.E., 1989, ApJ 336, 185

\bibitem[\protect\astroncite{Peterson et~al.}{1995}]{prc:95}
Peterson R.C., Rood R.T., Crocker D.A., 1995, ApJ 453, 214

\bibitem[\protect\astroncite{Pilachowski et~al.}{1996}]{Pilachowski96}
Pilachowski C.A., Sneden C., Kraft R.P., Langer G.E., 1996, AJ 112, 545

\bibitem[\protect\astroncite{Powell}{1999}]{Pow:99}
Powell D.C., 1999, PASP 111, 1186

\bibitem[\protect\astroncite{Reimers}{1975}]{r:75}
Reimers D., 1975, Mem. Soc. Roy. Sci. Li\`ege 8, 369

\bibitem[\protect\astroncite{Shetrone}{1996{a}}]{shet:96a}
Shetrone M.D., 1996{a}, AJ 112, 1517

\bibitem[\protect\astroncite{Shetrone}{1996{b}}]{shet:96b}
Shetrone M.D., 1996{b}, AJ 112, 2639

\bibitem[\protect\astroncite{Smith \& Tout}{1992}]{st:92}
Smith G.H., Tout C.A., 1992, MNRAS 256, 449

\bibitem[\protect\astroncite{Sneden et~al.}{1997}]{sks:97}
Sneden C., Kraft R.P., Shetrone M.D. et~al., 1997, AJ 114, 1964

\bibitem[\protect\astroncite{Sneden et~al.}{1991}]{sneden91}
Sneden C., Kraft R.P., Prosser C.F., Langer G.E., 1991, AJ 102, 2001

\bibitem[\protect\astroncite{Sweigart}{1997{a}}]{swei:97}
Sweigart A.V., 1997{a}, ApJL 474, L23

\bibitem[\protect\astroncite{Sweigart}{1997{b}}]{sweip:97}
Sweigart A.V., 1997{b}, in A.~G.~D. Philip, J. Liebert, R. Saffer, and D.~S.
  Hayes (eds.), The Third Conference on Faint Blue Stars, no.~95 in IAU Colloq.
  L. Davis Press

\bibitem[\protect\astroncite{Sweigart \& Mengel}{1979}]{sm:79}
Sweigart A.V., Mengel K.G., 1979, ApJ 229, 624

\bibitem[\protect\astroncite{VandenBerg et~al.}{1998}]{vlp:98}
VandenBerg D.A., Larson A.M., {De~Propris} R., 1998, PASP 110, 98

\bibitem[\protect\astroncite{Whitney et~al.}{1994}]{wor:94}
Whitney J.H., O'Connel R.W., Wood R.T., 1994, AJ 108, 1350

\bibitem[\protect\astroncite{Zahn}{1992}]{zahn:92}
Zahn J.P., 1992, A\&A 265, 115

\bibitem[\protect\astroncite{Zaidins \& Langer}{1997}]{zl:97}
Zaidins C., Langer G.E., 1997, PASP 109, 244

\end{thebibliography}

\end{document}